\documentclass[aps,prl,floats,twocolumn,twoside,floatfix]{revtex4}
\usepackage{amsmath}
\usepackage{graphicx}

\begin{document}
\title{Electronic correlation in cyclic polyenes.
Behavior of approximate coupled-pair theories for large rings}
\author{Rafa{\l} Podeszwa}
\affiliation{Department of Chemistry, University of Warsaw, Pasteura 1, PL-02-093 Warsaw, Poland}

\begin{abstract}
We investigate the cyclic polyenes (annulenes) ${\rm C}_M{\rm H}_M$, described
by the Pariser--Parr--Pople (PPP) model, by means of the approximate
coupled-pair theories (ACP, ACPQ).
For the systems with the spectroscopic value of the PPP resonance 
integral $\beta=2.5 \mbox{ eV}$,
the ACP method breaks down for $M\geq 446$ and the ACPQ
method---for $M>194$. 
In the ACPQ method, for $M>170$, two close
lying solutions have been observed that
become quasi-degenerate for $M \geq 198$.
The results indicate that the ACP and ACPQ methods
cannot be applied to the one-dimensional metallic case.
\end{abstract}
\maketitle
\section{Introduction}

The problem of electronic correlation in 
the $\pi$-electron systems corresponding to
small cyclic polyenes (annulenes)
has been studied extensively within the Pariser--Parr-Pople (PPP) model by 
Paldus {\em et al.}~\cite{paldus:84a,paldus:84b,paldus:82,takahashi:83,paldus:84c,takahashi:85,piecuch:90a,piecuch:90b,piecuch:90c,piecuch:91,paldus:92,piecuch:92}.
They have found that the coupled-cluster doubles method (CCD) [in annulenes
equivalent to the coupled-cluster singles and  doubles (CCSD) one]
breaks down when the correlation
effects become sufficiently strong.
They devised the approximate coupled-pair method
corrected for connected quadruple
excited clusters (ACPQ) \cite{paldus:84a,paldus:84b}. 
The ACPQ method, together with the similar ACP-D45 
method (or ACP, in short)~\cite{jankowski:80}, was shown to perform very well for the small 
annulenes, being convergent and giving correlation
energies close to the full configuration interaction (FCI) results. 

Recently, small annulenes have been studied
with the inclusion of the doubles (D), triples (T), and quadruples (Q)
in the cluster operator~\cite{podeszwa:02a}. It turned
out that although the inclusion of quadruple
excitations considerably improves the results 
in the weakly and moderately correlated regime (which correspond
to large and moderate absolute values of the PPP resonance integral
$\beta$, respectively), the CCDT, CCDQ, and CCDTQ methods break down 
in the strongly correlated regime.
It was also found that in the strongly correlated regime 
the $t_2$ amplitudes of the ACP and 
ACPQ methods differ considerably from the FCI $t_2$ amplitudes.

Even though the ACP and ACPQ methods are
satisfactory for small annulenes, it has been unknown
how the methods perform for large
annulenes, where the quasi-degeneracy problems
may be too strong. 
The strength of the
correlation (and the level of quasi-degeneracy) can be adjusted 
by decreasing the absolute value
of $\beta$ for small annulenes, but
it has been unknown whether this procedure is equivalent
to the increasing of the size of the ring and
keeping the resonance integral constant.
The behavior of the methods
for large systems is important if one wants to extrapolate
the results and study the limit of the one-dimensional metal.
That motivated us to investigate the performance
of the ACP and ACPQ methods for large annulenes.

\section{Theory}
In this Letter, we provide only a brief description
of the PPP model of annulenes; more details
may be found in Ref.~\citealp{podeszwa:02a}.

We shall consider annulenes of formula
${\rm C}_M{\rm H}_M$, where $M=4m_0+2$, $m_0=1,2,\ldots {}$,
and the number of electrons $N=M$.
The C atoms form a regular polygon, and the C--C bonds are 
assumed to be of the length $R^0 = 1.4$~\AA. 
The Fock-space Hamiltonian $\hat{H}$, built according to the 
prescriptions of the PPP model, is given in Eq.~(2) of Ref.~\citealp{podeszwa:02a}.
The following semiempirical parameters are used: the Coulomb integral $\alpha=0$,
the resonance integral $\beta= -2.5$~eV (the so-called spectroscopic value),
and the two-center two-electron integrals
$\gamma_{mn}$ are parametrized with the Mataga--Nishimoto formula~\cite{mataga:57},
$\gamma(R) = e^2 [R + e^2 (\gamma^{\rm 0})^{-1}]^{-1} \,$, where $e$ is the electron charge and $\gamma^0 = \gamma(0) = 10.84$~eV.
Within the PPP computer code, the atomic units are used:
$1 \mbox{ bohr}=0.529177\mbox{ \AA}$, $1 \mbox{ hartree}=27.2116\mbox{ eV}$
(the conversion factors are the same as employed in 
Refs.~\citealp{stolarczyk:88} and \citealp{podeszwa:02a}).

The ACP and ACPQ methods are approximations
to the CCD method.
In this method we use exponential Ansatz
and expand the ground-state wave function $\Psi$
by using the cluster operator $\hat{T}=\hat{T}_2$,
\begin{equation}
\Psi=\exp(\hat{T_2}) \Phi,
\end{equation}
where $\Phi$ is a single-determinantal reference
configuration, usually the RHF wave function.
The $\hat{T_2}$ operator depends linearly
on some parameters, the so-called $t_2$ amplitudes,
that can be obtained by solving the set of
nonlinear CCD equations given by
\begin{equation}
\label{eq:CCD_general}
\langle \Phi^{*} |\exp(-\hat{T_2})\hat{H}\exp(\hat{T}_2)\Phi\rangle = 0,
\end{equation}
where $\Phi^{*}$ 
represent all the doubly excited configurations.
The correlation energy for the ground state
is completely determined by the $t_2$ amplitudes
and can be calculated as
\begin{equation}
E_{\rm corr}=\langle \Phi |\hat{H}\hat{T}_2\Phi\rangle - E_{\rm HF}.
\end{equation}

The cyclic symmetry of our PPP model
imposes some extra conditions on the CC model.
The $t_1$ amplitudes (singles) must vanish,
and therefore $\Phi$ is also the Brueckner determinantal 
function~\cite{paldus:73,stolarczyk:84}.
Thus, the CCD method becomes equivalent to the CCSD one.
In the presence of the cyclic symmetry, the amplitudes
depend on 3 indices (instead of 4), and the number of amplitudes scales as $M^3$
(instead of $M^4$).
Moreover, the computational cost for the CCD iteration
scales as $M^4$ instead of $M^6$. This makes
enormous savings in the CPU time for large annulenes
and makes them computationally accessible.
The ACP and ACPQ methods violate the alternancy
symmetry~\cite{koutecky:85}, and therefore
we have not used this symmetry in the present calculations.
The details of the implementation of the cyclic symmetry
are given in Ref.~\citealp{podeszwa:02a}.

\begin{figure}[tb]
\begin{center}
\includegraphics[scale=1.0]{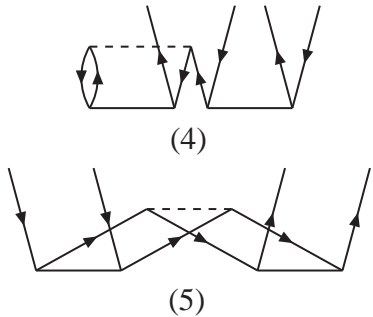}
\end{center}
\caption{The Brandow diagrams representing the quadratic terms in the ACP method.}
\label{acp-diagrams}
\end{figure}

The explicit form of the CCD equations
is most conveniently presented in the form of diagrams
(see, e.g., Ref.~\citealp{kucharski:86}). The left hand side of Eq.~(\ref{eq:CCD_general})
is then a sum of terms represented by these diagrams.
In the ACP method, some of the quadratic
terms of the CCD equations are omitted
(these are the diagrams that do not 
factorize with respect to the hole lines).
It was shown~\cite{paldus:84b} that these terms are 
approximately canceled by the terms
corresponding to the connected quadruple
excitations. Thus, the omission of these terms
may improve the results and was
shown to be very effective
for the annulenes in the strongly correlated regime~\cite{paldus:84b}.
The only quadratic terms that are present in the ACP
method  correspond to the diagrams in Fig.~\ref{acp-diagrams} (the labels below are the same as in Ref.~\citealp{paldus:84a},
where the complete set of the diagrams may be found).

For closed-shell calculations, the CC equations may be put in a
spin-adapted form. 
We use the nonorthogonally spin-adapted formalism~\cite{cizek:66},
which for the CCD method and its variants is equivalent
to the orthogonally spin-adapted formalism used
by Paldus {\em et al.}~\cite{paldus:84b}.
With the latter formalism they
devised an improvement to the ACP method called ACPQ.
In this method the diagram~(5) of Fig.~\ref{acp-diagrams}
is multiplied by~9 in the triplet coupled equations of Ref.~\citealp{paldus:84b}.
The ACPQ method better approximates the quadruples
and was shown to provide the exact solution in the limit of 
$\beta \to 0$~\cite{paldus:84b}.

The ACPQ method can also be translated to the
nonorthogonally spin-adapted formalism.
Let $g_{ij}^{ab}$ represent the 
nonorthogonally spin-adapted term corresponding to 
diagram (5) of Fig.~\ref{acp-diagrams}
in the ACP method ($i,j$ and $a,b$ stand for the occupied and unoccupied orbitals,
respectively). In the ACPQ method, 
the term should be modified in the following way:
\begin{equation}
(\mbox{ACP})  \quad g_{ij}^{ab}  \to 5g_{ij}^{ab}-4g_{ji}^{ab} \quad (\mbox{ACPQ})
\end{equation}

\begin{figure}[t]
\begin{center}
\includegraphics[scale=1.0]{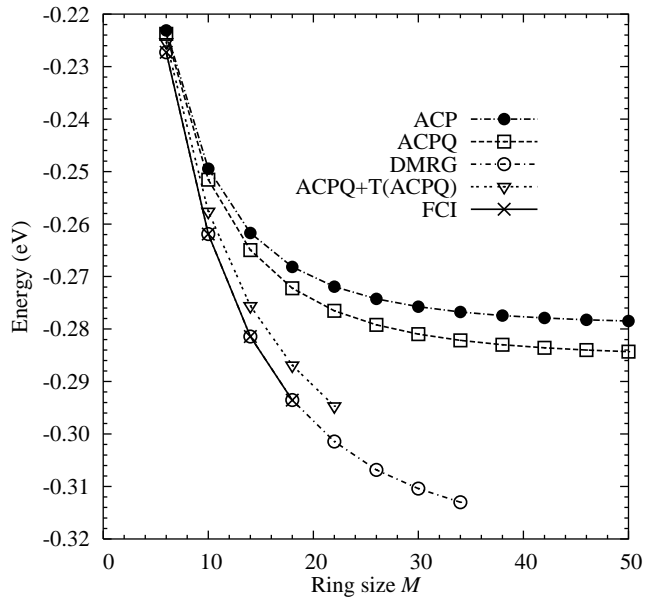}
\end{center}
\caption{The correlation energy per electron (in eV) for small annulenes.}
\label{fig:male}
\end{figure}

\section{Results}
We performed the ACP and ACPQ calculations
for the annulenes, ${\rm C}_M{\rm H}_M$, for $M$
up to $442$. In Fig.~\ref{fig:male},
the ACP and ACPQ results for small annulenes are compared with the 
results of ACPQ+T(ACPQ)~\cite{paldus:92}, FCI results~\cite{bendazzoli:94}, and 
the density matrix renormalization group (DMRG)
calculations~\cite{fano:98}.
These are the methods that were shown to
provide converged results for the small annulenes.
It can be seen that the correlation energy
corrected for the triples 
in the ACPQ+T(ACPQ) method is much better than in the ACPQ method.
The method depends, however, on the $t_2$ amplitudes from the ACPQ method,
and will work only if the latter method converges.
For small annulenes the DMRG results are the most accurate
and the deviation from the FCI is so small that it cannot
be seen on the figure. 
 
\begin{figure}[tb]
\begin{center}
\includegraphics[scale=1.0]{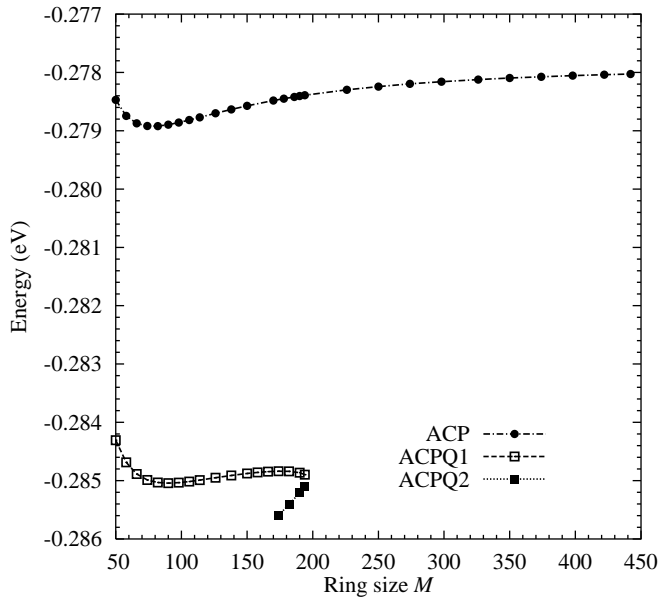}
\end{center}
\caption{The ACP and ACPQ correlation energy per electron (in eV) for large annulenes.
}
\label{fig:duze}
\end{figure}

Since the ACP and ACPQ correlation energies do not change
much with the increasing size of the ring around $M=50$, 
it may suggest that the results are close
to the saturation. However, it is not true for the larger annulenes,
as shown in Fig.~\ref{fig:duze}. Some numerical values
are also shown in Table~\ref{tab:duze}.
Surprisingly, the
correlation energy is not monotonic with the increasing size of the ring
and reaches a minimum. It means that the extrapolation of the 
results for smaller rings would lead to a completely wrong result.

\begin{table*}[tb]
\caption{The correlation energy per electron (in eV) for large annulenes; 
NC means no convergence of the DIIS iterations, 
NF means that the result was not found}
\label{tab:duze}
\begin{center}
\begin{tabular}{rrrrrr}
\hline \hline
  & \multicolumn{1}{c}{50}& \multicolumn{1}{c}{98}& \multicolumn{1}{c}{174}& \multicolumn{1}{c}{194}& \multicolumn{1}{c}{442}\\
\hline
ACP& $-0.278\mbox{ }470$& $-0.278\mbox{ }859$& $-0.278\mbox{ }466$& $-0.278\mbox{ }392$& $-0.278\mbox{ }028$\\
ACPQ1& $-0.284\mbox{ }306$& $-0.285\mbox{ }034$& $-0.284\mbox{ }838$& $-0.284\mbox{ }898$&\multicolumn{1}{c}{NC}\\
ACPQ2&\multicolumn{1}{c}{NF}&\multicolumn{1}{c}{NF}& $-0.285\mbox{ }599$& $-0.285\mbox{ }096$&\multicolumn{1}{c}{NC}\\
\hline \hline
\end{tabular}
\end{center}
\end{table*}

We found that the ACP method is convergent up to $M=442$.
For $M>270$, the use of the direct inversion in the iterative
subspace method
(DIIS)~\cite{pulay:82,scuseria:86}
was necessary to provide converged results. The DIIS
method was also very useful in accelerating
the convergence for the smaller annulenes.

The results of the ACPQ method are even more complicated.
For $M<174$ there was one solution
found, but between 174 and 194 we found two solutions.
One set of solutions, later referred to as ACPQ1, is
a continuation of the results for $M<174$.
It was found by using the ACP result for a given $M$ as a starting point 
for the iterative process with the DIIS method.
The other set of solutions, of lower energy, was found by applying
the DIIS iterative scheme to the MP2 starting point. For $M<174$,
both starting points 
resulted in the same solution (ACPQ1).

\begin{figure}[tb]
\begin{center}
\includegraphics[scale=1.0]{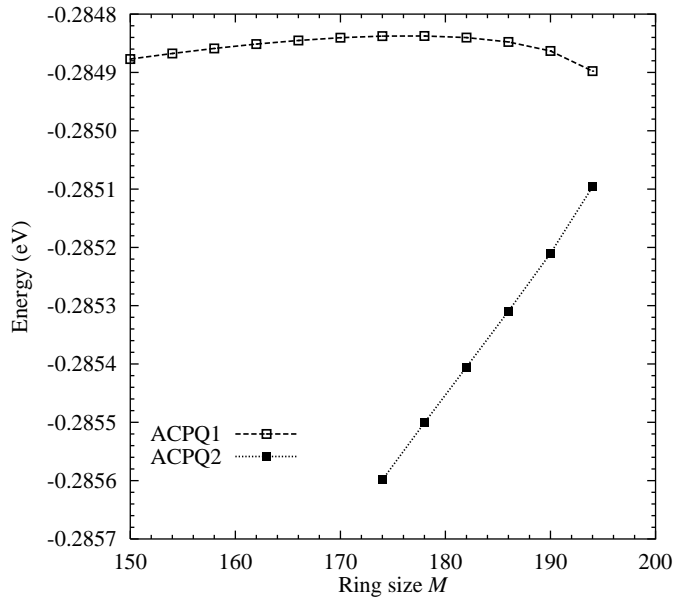}
\end{center}
\caption{The ACPQ correlation energy per electron (in eV) for large annulenes.
}
\label{fig:acpq}
\end{figure}

None of the ACPQ1 and ACPQ2 converged for
$M\geq 198$. In Fig.~\ref{fig:acpq}, the results
of the ACPQ method for $M$ from $150$ to $198$ are shown.
It can be seen that the correlation energies
of the two solutions approach each other
and the states become
quasi-degenerate for $M\to 198$.
It is interesting to check whether
the amplitudes of the states also approach each other.
To test the similarity
of the $t_2$ amplitudes of the two solutions, we 
calculated the parameters $\theta$ and $\eta$ defined 
in Eq.~(22) of Ref.~\citealp{podeszwa:02a}.
Here $\theta$ measures the angle between the vectors
formed of the $t_2$ amplitudes, 
and $\eta$---the ratio of the vector lengths.
For $M=174$, we found $\theta=2.69^{\circ}$ and $\eta=0.9989$, while
for $M=194$, $\theta=0.74^{\circ}$ and $\eta=0.9999$.
This shows that the $t_2$ amplitudes of the ACPQ1 and the ACPQ2
solutions are quite similar for $M=174$, and are almost
identical for $M=194$. 

The existence of multiple solutions of the CC
equations has been observed previously for the H4 model~\cite{meissner:93,jankowski:94},
and the complete set of solutions has been obtained with the
homotopy method~\cite{kowalski:98a}. In those studies, one of the
solutions was easily identified as the ground-state solution.
In contrast, for the ACPQ method of our annulene,
such identification is ambiguous.
The ACPQ1 solution, which exists for small annulenes,
and which may be considered as the standard solution,
has the energy {\em higher\/} than the ACPQ2 solution. 
Some exotic solutions that have the energy lower than
the ground state have been observed for the PPP model of benzene~\cite{podeszwa:02b}.
Our preliminary results for $\rm C_{10} H_{10}$  also
show the existence of such solutions~\cite{podeszwa:02d}.
 
\section{Conclusions}

The ACP an ACPQ methods, while performing
satisfactorily for small annulenes, have 
convergence problems for large annulenes.
The ACP and ACPQ correlation energies are not monotonic with the increasing size of the ring.
It shows that an extreme caution must be taken
if one tries to extrapolate
the results of the small annulenes 
to the infinite limit. 
In the case of the ACPQ method, two solutions are found for $174\leq M \leq 194$,
behaving as if they coalesce into a single solution between $M=194$ and $M=198$.
However, no converged real solution of the ACPQ equations
has been found beyond $M=194$,
which may indicate that the solutions become complex
for $M\geq198$.
The above observations indicate that
the ACP and ACPQ methods
cannot be applied to the one-dimensional metallic case.

Among all the methods
used in the calculations of the PPP annulenes,
the DMRG method seems the most promising. However,
it should be noted that the
method was not tested in a broader range of 
the resonance integral $\beta$, especially in the strongly
correlated regime of $\beta\approx 0$, where the various CC
methods break down. It is also unclear whether the method
can be applied to the large or infinite annulenes,
since the method may not be size-extensive
for the metallic case~\cite{chan:02}.

The existence of a method that can
describe the large annulenes and the 
one-dimensional metal in the PPP model is still an open problem.
It is unlikely that the method that
cannot describe the strongly correlated regime of
a small annulene would be able to describe the 
infinite case. It seems, however, that the proper behavior of a method
in the strongly correlated regime of small annulenes
still does not guarantee the correct description of the
larger systems.

I would like to express my gratitude to L. Z. Stolarczyk
for encouragement and helpful discussions. I would also like to thank 
B. Jeziorski and L. Z. Stolarczyk for the critical reading of the
manuscript.
The work was supported by the Committee for Scientific Research 
(KBN) through Grant No. 7 T09A  019 20.


\end{document}